\documentclass[letter]{aa} 
\usepackage{graphicx}
\usepackage{txfonts}
\begin{document}
   \title{Star-forming galaxies in SDSS: signs of metallicity evolution}

   \author{M.A. Lara-L\'opez
          \inst{1}
          \and
          J. Cepa\inst{1,2}
	  \and
          A. Bongiovanni\inst{1}
	  \and
          H. Casta\~neda\inst{1}
	  \and
	  A.M. P\'erez Garc\'{\i}a\inst{1}
	  \and
          M. Fern\'andez Lorenzo\inst{1}
	  \and
          M. P\'ovic\inst{1}
	  \and
          M. S\'anchez-Portal\inst{3}
         }

   \institute{Instituto de Astrof\'{\i}sica de Canarias, 38200 La Laguna, Spain
              \email{mall@iac.es}
         \and
             Departamento de Astrof\'{\i}sica, Universidad de la Laguna, Spain
          \and
             Herschel Science Center, INSA/ESAC, Madrid, Spain
             }

   \date{Received; accepted}

\abstract  
{Evolution of galaxies through cosmic time has been widely studied at high redshift, but there are a few studies in this field at lower redshifts. However, low--redshifts studies will provide important clues to the evolution of galaxies, furnishing the required link between local and high--redshift universe.} 
{In this work we focus on the metallicity of the gas in spiral galaxies at low redshift looking for signs of chemical evolution. We analyze the metallicity contents of star forming galaxies of similar luminosities at different redshifts, we studied the metallicity of star forming galaxies from SDSS--DR5 (Sloan Digital Sky Survey--Data Release 5), using different redshift intervals from 0.1 to 0.4.}
{We used the public data of SDSS--DR5 processed with the STARLIGHT spectral synthesis code, correcting the fluxes for dust extinction, estimating metallicities using the $R_{23}$ method, and analyzing the samples with respect to the [{N\,\textsc{ii}}] $\lambda$6583/[{O\,\textsc{ii}}] $\lambda$3727 line ratio.}
{From a final sample of 207 galaxies, we find a decrement in 12+log(O/H) corresponding to the redshift interval 0.3 $<$ z $<$ 0.4 of $\sim$0.1 dex with respect to the rest of the sample, which can be interpreted as evidence of the metallicity evolution in low--z galaxies.}
{}

   \maketitle
%

\section{Introduction}

Determination of the chemical composition of the gas in galaxies versus cosmic time provides a very important tool for understanding galaxy evolution, due to its important impact on fields such as stellar evolution and nucleosynthesis, gas enrichment processes, and the primary or secondary nature of the different species.

Optical emission lines in galaxies have been widely used to estimate abundances in extragalactic {H\,\textsc{ii}} regions (e.g. Aller 1942, Searle 1971, Pagel 1986, Shields 1990). The direct method of estimating metallicity in galaxies is known as the $``$$T_e$ method$"$ (Pagel et al. 1992, Skillman $\&$ Kennicutt 1993), and consists of measuring the oxygen abundance from the electron temperature of the {H\,\textsc{ii}} region obtained using, the ratio of a high--excitation auroral line such as [{O\,\textsc{iii}}] $\lambda$4363 to the lower excitation [{O\,\textsc{iii}}] $\lambda\lambda$ 4959, 5007 lines.

However, [{O\,\textsc{iii}}] $\lambda$4363 is too weak, not only in metal--rich, but even in metal--poor galaxies. In the absence of this auroral line, metallicities have to be estimated from strong line ratios, such as the $R_{23}$ ratio introduced by Pagel et al. (1979). This is a commonly used and well--calibrated method of estimating metallicity abundances (see Alloin et al. 1979, Edmunds $\&$ Pagel 1984, McCall et al. 1985, Dopita $\&$ Evans 1986, Pilyugin 2000, 2001, Tremonti et al. 2004, Kewley $\&$ Dopita 2002, Liang et al. 2006, hereafter L06). Nevertheless, this method has the disadvantage of being double--valued as a function of 12$+$log(O/H). Besides  $R_{23}$, there are other strong--line methods useful for determining abundances of high--metallicity, star--forming galaxies, for example the N2=[{N\,\textsc{ii}}] $\lambda$6583/ {H$\alpha$} method (Denicol\'o et al. 2002, Pettini $\&$ Pagel 2004), the  $S_{23}$=([{S\,\textsc{ii}}] $\lambda\lambda$6717, 6731 +[{S\,\textsc{iii}}] $\lambda\lambda$9069,9532)/{H$\beta$} (V\'{\i}lchez  $\&$  Esteban 1996), among others. For a detailed discussion of the different methods see Bresolin (2006) and Kewley  $\&$  Ellison (2008).

Many studies of metallicity evolution exist as a function of cosmic time,  although many of them refer to damped Lyman $\alpha$ systems at z $>$ 2 (e.g. Pettini et al. 2002, Henry $\&$ Prochaska 2007). For the evolution of the mass--metallicity relation of star--forming galaxies, Erb et al. (2006) find evolution at z $\gtrsim$ 2 and z $\sim$ 3.5,  respectively. Also, Brooks et al. (2007) and Finlator $\&$ Dav\'e (2008), among others, derive cosmological models of the mass--metallicity relation, while at intermediate redshifts (1 $<$ z $<$ 2),  there are several important studies of the evolution of the chemical composition of the gas, such as the ones by Lilly et al. (2003), Kobulnicky et al (2003), and Maier et al. (2006).

Among the studies at z $<$ 1, mainly based on small samples,  Savaglio et al. (2005) have investigated the mass--metallicity relations using galaxies at 0.4 $<$ z $<$ 1, finding that the metallicity is lower at higher redshift, for the same stellar mass, by ~0.15 dex. Also, Maier at al. (2005), with a sample of 30 galaxies with 0.47 $<$ z $<$ 0.92, find that one--third have  metallicities lower than those of local galaxies with similar luminosities and star formation rates. In contrast, Carollo $\&$ Lilly (2001) use emission--line ratios of 15 galaxies in a range of 0.5 $<$ z $<$ 1 to find that their metallicities appear to be remarkably similar to those of local galaxies selected with the same criteria. A similar result was found for the luminosity--metallicity relation by Lamareille et al. (2006), comparing star--forming galaxies at local and intermediate redshift (0.2 $<$ z $<$ 1, split into 0.2 redshift bins). However, Buat et al. (2008) and Kobayashi et al. (2007) have derived a model of metallicity evolution as a function of $z$, which show a progressive increase in metallicity with time, even at low redshift.

There is, then, a need to increase the lower redshift galaxy samples to ascertain whether or not at such low redshifts (i.e. small lookback times, of the order of 8.4 Gyr for z $\sim$ 0.5, using a concordance cosmology) any evidence exists of metallicity evolution. The SDSS database provides an excellent opportunity for extending these studies up to z $\sim$0.4 to explore a possible evolution of metallicity at low--redshift using larger samples, thus deriving more statistically significant results. 


\section{Sample selection}

We used the SDSS--DR5 (Adelman--McCarthy et al. 2007) spectra from the STARLIGHT Database\footnote{http://www.starlight.ufsc.br}, which were processed through the STARLIGHT spectral synthesis code developed by Cid Fernandes and colleagues (Cid Fernandes et al. 2005, 2007, Mateus et al. 2006, Asari et al. 2007). We obtained the emission lines fluxes measurements of our samples from the continuum subtracted  spectra. For each emission line, STARLIGHT code returns the rest frame flux and its associated equivalent width (EW), linewidth, velocity displacement relative to the rest--frame wavelength and the S/N of the fit. In the case of Balmer lines, the underlying stellar absorption was corrected using synthetic spectra obtained by fitting an observed spectrum  with a combination of $N_{\star}$=150  simple stellar population (SSPs) from the evolutionary synthesis models of Bruzual $\&$ Charlot (2003). 

Since we are interested in studying the chemical evolution of emission--line galaxies at different redshifts, our sample was divided into three intervals: $z_1$=(0.1$-$0.2), $z_2$=(0.2$-$0.3), and $z_3$=(0.3$-$0.4). To avoid luminosity biases, we selected galaxies in the same luminosity intervals, taking as a reference the magnitude completeness of the farthest interval $z_3$. Then, we selected sample galaxies with absolute Petrosian r magnitude in the range $-$24.8 $<$ $M_r$ $<$ $-$23.1 mag. As a consequence, we end with a sample spanning a narrow luminosity range, thus avoiding luminosity--metallicity evolutionary effects, as will be shown in Fig. 5. Absolute magnitudes were both, K  and Galactic extinction corrected by using the code provided by Blanton et al. (2003) and the maps of Schlegel et al. (1998), respectively, as provided by the STARLIGHT team. The Schlegel resulting sample contains 25812 galaxies for $z_3$, 15193 for $z_2$ and 4225 for $z_1$. It was not possible to extend the sample to redshifts lower than 0.1 due to the scarce number of spectra.

From this sample we only consider galaxies whose spectra show in emission the {H$\alpha$}, {H$\beta$}, [{N\,\textsc{ii}}] $\lambda$6583, [{O\,\textsc{ii}}] $\lambda$3727, [{O\,\textsc{iii}}] $\lambda$4959, and [{O\,\textsc{iii}}] $\lambda$5007 lines, with a signal--to--noise ratio for [{O\,\textsc{ii}}] $\lambda$3727 higher than 3$\sigma$. A final selection for star--forming galaxies excluding AGNs was made using the criteria given by Kauffmann et al. (2003) in the [{O\,\textsc{iii}}] $\lambda$5007/{H$\beta$} vs. [{N\,\textsc{ii}}] $\lambda$6583/{H$\alpha$} diagram, leaving a sample of 34 galaxies for $z_3$, 142 for $z_2$, and 40 for $z_1$.

Since Balmer lines have already been corrected for underlying stellar absorption by the STARLIGHT code,  it is only necessary to correct emission lines for dust extinction. Our extinction correction was derived using the Cardelli extinction curve (Cardelli et al. 1989) and the Balmer decrements, assuming case B recombination for a density of 100 cm$^{-3}$ and a temperature of 10$^{4}$ K, results in a  {H$\alpha$}/{H$\beta$} predicted ratio (unaffected by reddening or absorption) of 2.86 (Osterbrock 1989).

\section{Metallicity estimates} 

We estimate metallicities using the $R_{23}$ relation, introduced by Pagel et al. (1979), $R_{23}$=([{O\,\textsc{ii}}] $\lambda$3727]+[{O\,\textsc{iii}}] $\lambda\lambda$4959, 5007])/H$\beta$, adopting the calibration given by Tremonti et al. (2004), 12+log(O/H) = 9.185 $-$ 0.313$x$ $-$ 0.264$x^2$ $-$ 0.321$x^3$, where $x$ = log $R_{23}$. 

However, this calibration is valid only for the upper branch of the double--valued $R_{23}$ abundance relation, and additional line ratios, such as [{N\,\textsc{ii}}] $\lambda$6583/[{O\,\textsc{ii}}]$\lambda$3727, are required to break this degeneracy. Since the upper and lower branches of the $R_{23}$ calibration bifurcates at log([{N\,\textsc{ii}}]/[{O\,\textsc{ii}}]) $\sim$ --1.2 for the SDSS galaxies (Kewley $\&$ Ellison 2008), which corresponds to a metallicity of 12+log(O/H) $\sim$ 8.4, we select galaxies having  12$+$log(O/H) $>$ 8.4 and log([{N\,\textsc{ii}}]/[{O\,\textsc{ii}}]) $>$ --1.2, corresponding to the upper $R_{23}$ branch. Applying this criterion, we end up with a final sample of 28 galaxies for $z_3$, 140 for $z_2$, and 39 for $z_1$, by discarding  galaxies of the lower branch we are not introducing a bias because 96$\%$ of our galaxies lie in the upper branch.

From these data, we derived the abundance--sensitive diagnostic diagram [{N\,\textsc{ii}}] $\lambda$6583/[{O\,\textsc{ii}}] $\lambda$3727 vs. 12+log(O/H), represented in Fig. 1. We selected this diagram due to its low scatter and physical information. The advantages of using [{N\,\textsc{ii}}] and [{O\,\textsc{ii}}] lines is that they are not affected by underlying stellar population absorption, and because this ratio is almost independent of the ionization parameter, since $\rm{N}^{+}$ and ${\rm{O}}^{+}$ have similar ionization potentials.

\begin{figure}[htp]
   \centering
   \includegraphics[scale=0.50]{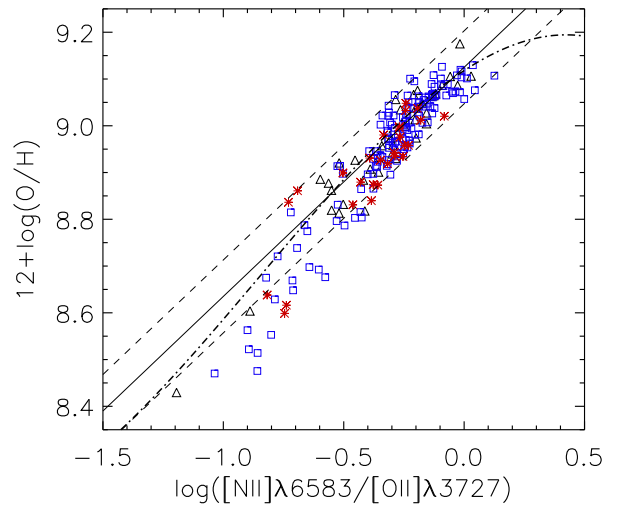}
      \caption{Calibration relation between 12+log(O/H) and log([{N\,\textsc{ii}}] $\lambda$6583/[{O\,\textsc{ii}}] $\lambda$3727) for the three samples in redshift, triangles represent galaxies of the sample at redshift $z_1$, squares galaxies of $z_2$, and asterisks galaxies of $z_3$. The solid line represents the best fit of the data using a linear fit, with the 2$\sigma$ discrepancy in short dashed lines, and the dot--dash line for an  order three polynomial fit given by L06}
         \label{FigVibStab}
   \end{figure}
   
This diagram has also been used by Kewley $\&$ Dopita (2002), Nagao et al. (2006), and L06 among other metallicity--sensitive emission--line ratios, like log([{N\,\textsc{ii}}] $\lambda$6583/H$\alpha$), log([{O\,\textsc{iii}}] $\lambda$5007/H$\beta$)/[{N\,\textsc{ii}}] $\lambda$6583/H$\alpha$), and log([{O\,\textsc{iii}}] $\lambda$4959, 5007/H$\beta$). Figure 1 shows our three samples in redshift, as well as the fits of L06 for their sample (0.04 $<$ $z$ $<$ 0.25, and Petrosian $r$ magnitude in the range 14.5 $<$ $m_r$ $<$ 17.77).
   
To interpret our results, it is important to note that we are working with the integrated spectra, but that we are comparing with previous studies of nuclear spectra. As shown by Kewley et al. (2005) and L06, using a sample of Jansen et al. (2000), data points from nuclear spectra follow the SDSS galaxies nuclear spectra very well, but data points from the integrated spectra show lower 12+log(O/H). The nuclear metallicities exceed the metallicities derived from integrated spectra by $\sim$0.13 dex on average.

Based on this diagram, we claim that an evolution with redshift is present in our sample . There is a clear decrement of $\sim$0.07 dex in 12+log(O/H), corresponding to the samples of $z_1$ and $z_2$ compared to the L06 polynomial fit, an effect resulting from the fiber size with respect to the size of the galaxies, as pointed out by Kewley et al. (2005). Both samples are also $\sim$0.4 dex lower in  [{N\,\textsc{ii}}] $\lambda$6583/[{O\,\textsc{ii}}] $\lambda$3727. However, the $z_3$ sample is $\sim$0.1 dex lower in 12+log(O/H) with respect to the $z_1$ and $z_2$ samples. Samples corresponding to $z_1$ and $z_2$, seem to follow the same distribution in metallicity, but the $z_3$ sample shows a clear decrement in 12+log(O/H) of $\sim$0.1 dex and [{N\,\textsc{ii}}] $\lambda$6583/[{O\,\textsc{ii}}] $\lambda$3727 of $\sim$0.2 dex related to the other two samples.

\section{Metallicity evolution}

To validate the decrement observed in Fig. 1, we proceed to generate a histogram of metallicities for our three redshifts samples as shown in Fig. 2. In the histograms we can appreciate a shift to lower metallicities as redshift increases, even for the $z_1$ and $z_3$ samples, in which we have only 39 and 28 galaxies, respectively.

\begin{figure}
\centering
\includegraphics[scale=0.50]{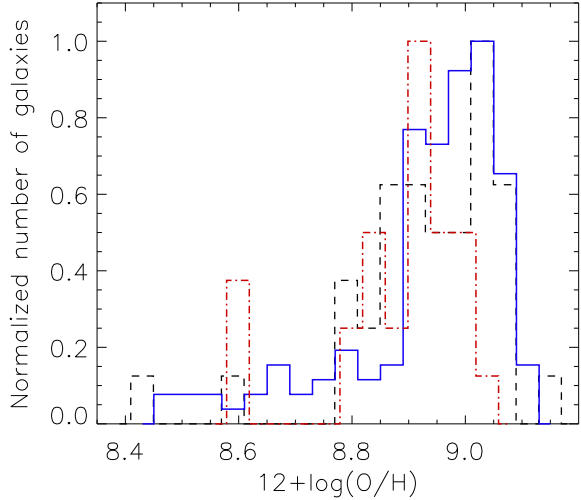}
\caption{Normalized metallicities histograms for our three samples. The dashed line represents the sample in the redshift interval $z_1$, the sample at $z_2$  is symbolized by a solid line, and the dot--dashed line corresponds to the sub--sample at redshift $z_3$ }
\label{FigVibStab}
\end{figure}

Although the derived distributions suggest a metallicity evolution, to investigate whether or not these could be an artifact due to the limited number of galaxies, we performed Monte Carlo simulations to determine the probability that the three distributions represent the same sample. Taking as a reference the $z_2$ distribution, because it has the larger number of galaxies, we simulated this distribution by fitting a Gaussian, but using the number of galaxies of the other two samples. After a thousand simulations, we find an 8$\%$ and a 3.5$\%$ probability that this distribution represents the same as that of $z_1$, and $z_3$, respectively. This result supports the idea of an intrinsic evolution in metallicity as observed at $z_3$.

\section{Discussion}

To investigate whether the origin of the decrement in metallicity of sample $z_3$ is due to instrumental effects or to an inherent property of the sample of galaxies, it is necessary to explore three important effects: the origin of nitrogen, the effect of the 3 arcsec diameter of the Sloan fibers, and the luminosity--metallicity relation. 

\subsection{Nitrogen and oxygen abundances}

The observed decrement in metallicity at the higher redshift could be due to the primary or secondary origin of nitrogen as redshift increases. To determine the origin of nitrogen in our samples, we estimate the abundance for these galaxies, a detailed explanation of this method will be given in our next paper (in preparation). We do not appreciate a significant difference between our redshift samples concerned with the primary or secondary origin of nitrogen. For the sample $z_3$, a 57.2$\%$ of the galaxies lie in the primary zone of the nitrogen, and the rest in the secondary part. Therefore, the decrement observed in metallicity is not due to a bias in the primary or secondary origin of nitrogen in the samples. Our results agree with those obtained by Vila--Costas $\&$ Edmunds (1993), L06, Kennicutt et al. (2003).

\subsection{Effect of the Sloan fiber diameter}

With respect to the Sloan fibers, we expect that at higher redshift, the projected size of the Sloan fibers will cover a larger fraction of galaxy area than for nearby galaxies. As argued in Sect. 3, since integrated metallicities are lower than nuclear, this could explain the metallicity decrease that can be observed for the $z_3$ sample with respect to the other ones. To quantify this effect for our sample, we estimated the percentage of angular size of each galaxy inside the Sloan fiber using the Petrosian total radius in r band in arcsec.

We find that the 3 arcsec diameter of the Sloan fibers cover a maximum of 50$\%$ of the angular size of the galaxies for the 100$\%$ of the $z_1$ sample, 98$\%$ of the $z_2$ sample, and 90$\%$ of the $z_3$ sample. It is clear that the differences are not significant and that the observed decrement in metallicity should not be attributed to a differential projected size of Sloan fibers at each sample.

\subsection{Effect of the luminosity--metallicity relation}

We also discard the effects of a luminosity--metallicity relation, because the galaxies in our sample do not show a dependence of luminosity versus metallicity, as show in Fig. 5, since we are using the same range of 1.7 absolute magnitudes at every redshift interval, which is too small to find significant differences in metallicity. Important variations in the luminosity--metallicity relation can only be seen  when spanning ranges of $\sim$5 magnitudes.\\

\begin{figure}
 \centering
\includegraphics[scale=0.50]{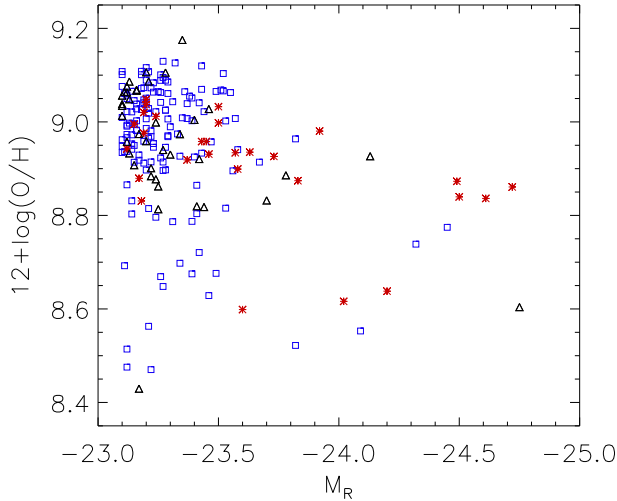}
\caption{Luminosity--Metalicity relation for our sample of galaxies, symbols follow the same code used in Fig. 1.}
 \label{FigVibStab}
 \end{figure}

\section{Conclusions}

Although similar studies of metallicity using SDSS exist, they are either restricted to a redshift $\sim$0.1 (e.g. Tremonti et al. 2004) or do not separate their samples as a function of redshift (e.g. L06), thus making it impossible to detect metallicity evolution. Finally, studies at high redshift are statistically limited. 

After exploring possible biases on the sample, the results obtained in the present work suggest  \emph{prima facie} evidence of an intrinsic metallicity evolution in the local Universe, showing a decrement of $\sim$0.1 dex in 12+log(O/H) at a redshift interval 0.3 $<$ z $<$ 0.4, a result consistent with the models of metallicity evolution in Buat et al. (2008) and Kobayashi et al. (2007).

\begin{acknowledgements}
This work was supported by the Spanish
\emph{Plan Nacional de Astronom\'{\i}a y Astrof\'{\i}sica} under grant AYA2005--04149. The Sloan Digital Sky Survey (SDSS) is a joint project of The University of Chicago, Fermilab, the Institute for Advanced Study, the Japan Participation Group, The Johns Hopkins University, the Max--Planck--Institute for Astronomy, Princeton University, the United States Naval Observatory, and the University of Washington. Apache Point Observatory, site of the SDSS, is operated by the Astrophysical Research Consortium. Funding for the project has been provided by the Alfred P. Sloan Foundation, the SDSS member institutions, the National Aeronautics and Space Administration, the National Science Foundation, the U.S. Department of Energy, and Monbusho. The official SDSS web site is www.sdss.org.  We thank the Starlight Project Team (UFSC, Brazil), especially William Schoenell, who helped us downloading the whole data set. We also thank the anonymous referee for his/her constructive comments.

\end{acknowledgements}

\end{document}